# IMPROVING COMPUTATIONAL EFFICIENCY OF MONTE-CARLO SIMULATIONS WITH VARIANCE REDUCTION

**A. Turner and A. Davis[*]**

EURATOM/CCFE Fusion Association
Culham Science Centre, Abingdon, Oxon, OX14 3DB, UK
andrew.turner@ccfe.ac.uk; davisa@engr.wisc.edu

## ABSTRACT

CCFE perform Monte-Carlo transport simulations on large and complex tokamak models such as ITER. Such simulations are challenging since streaming and deep penetration effects are equally important. In order to make such simulations tractable, both variance reduction (VR) techniques and parallel computing are used. It has been found that the application of VR techniques in such models significantly reduces the efficiency of parallel computation due to 'long histories'.

VR in MCNP can be accomplished using energy-dependent weight windows. The weight window represents an 'average behaviour' of particles, and large deviations in the arriving weight of a particle give rise to extreme amounts of splitting being performed and a long history. When running on parallel clusters, a long history can have a detrimental effect on the parallel efficiency – if one process is computing the long history, the other CPUs complete their batch of histories and wait idle. Furthermore some long histories have been found to be effectively intractable.

To combat this effect, CCFE has developed an adaptation of MCNP which dynamically adjusts the WW where a large weight deviation is encountered. The method effectively 'de-optimises' the WW, reducing the VR performance but this is offset by a significant increase in parallel efficiency. Testing with a simple geometry has shown the method does not bias the result. This 'long history method' has enabled CCFE to significantly improve the performance of MCNP calculations for ITER on parallel clusters, and will be beneficial for any geometry combining streaming and deep penetration effects.

## 1. INTRODUCTION

The Applied Radiation Physics group at Culham Centre for Fusion Energy (CCFE) supports the UK and international fusion research effort and performs nuclear analysis on large and complex models of magnetic confinement fusion devices (tokamaks) such as ITER and DEMO. Calculations are frequently performed using Monte-Carlo based radiation transport codes, for example the MCNP code [1]. The timely deliverable of results is critical, particularly for ITER related analysis, where results directly feed in to ongoing development and procurement programmes. Two acceleration techniques are employed at CCFE; variance reduction (VR), and parallel computing.

---

[*] New affiliation: University of Wisconsin-Madison, Madison, WI 53706, USA.

CCFE frequently perform 3-D activation calculations, which are important for planning the radiological safety and waste management aspects of future fusion devices. These calculations are performed using the in-house software MCR2S (Mesh-Coupled Rigerous-2-Step, [2]), and require the neutron flux and spectrum to be determined accurately obtained over a large spatial mesh, for which traditional VR techniques are often insufficient. To achieve this, so called 'global' VR (GVR) is used, and CCFE has developed a method of implementing MCNP weight windows (WW) to achieve accurate results over the entire model geometry.

In line with the method proposed by Cooper and Larsen [3], CCFE's GVR method utilises the forward flux solution to produce an estimate of the WW, however unlike [3] and other established deterministic methods such as FW-CADIS [4], the method does not require the production of an additional mesh-based geometry model, nor a transport calculation in a separate deterministic code. Instead, the forward flux solution is obtained over a mesh tally using a conventional MCNP calculation. The initial estimate of the forward flux will be poor in locations far from the source, however further iterations of this process using the previously generated WW map will progressively improve the result. The iterative process is continued until the flux is obtained throughout the problem space. Further details of the GVR method are given in [5].

It has been found that the use of strong VR techniques, particularly GVR, can lead to increased variation in the time taken to complete Monte-Carlo histories, which in turn reduces the effectiveness of parallelising the calculation. Extreme cases of such history length variation have been observed, so called 'long histories', which can reduce the effectiveness of parallel computing to the point of making a calculation intractable, which clearly undesirable.

## 2. LONG HISTORIES

Variance reduction in MCNP can be accomplished using weight windows (WW), where each region in phase space (space and energy) is assigned a target weight range. Particles with weight outside of this range are split or rouletted to maintain the required particle weight. If a particle were to find itself in a region of phase space with WW bounds orders of magnitude smaller than its own weight, it would be split into particles of smaller weight (typically a maximum of 5 by default). Assuming all the 'daughter' particles collide in the same region, these will in turn be split again, and so forth until all the daughter particles are within the WW bounds.

WW bounds are usually set according to some average quantity such as flux or track weight; however, infrequently sampled events can produce histories that deviate strongly from that average weight which leads to excess splitting. Such events may include a high weight particle entering a streaming path, and arriving in a region of much lower expected weight. Such long history effects have been observed by others [6], and are not solely confined to the above GVR implementation.

The presence of such 'long histories' (henceforth abbreviated to LH), have been observed when applying GVR techniques to MCNP model representations of the ITER device, such as the ITER reference model 'A-lite' shown in Figure 1 [7]. ITER containing a mixture of heavily shielded regions and small (< 1 cm) streaming paths, along with a spatially distributed neutron source.

Streaming paths and deep penetration effects are equally important in the ITER tokamak models, which seems particularly susceptible to LHs.

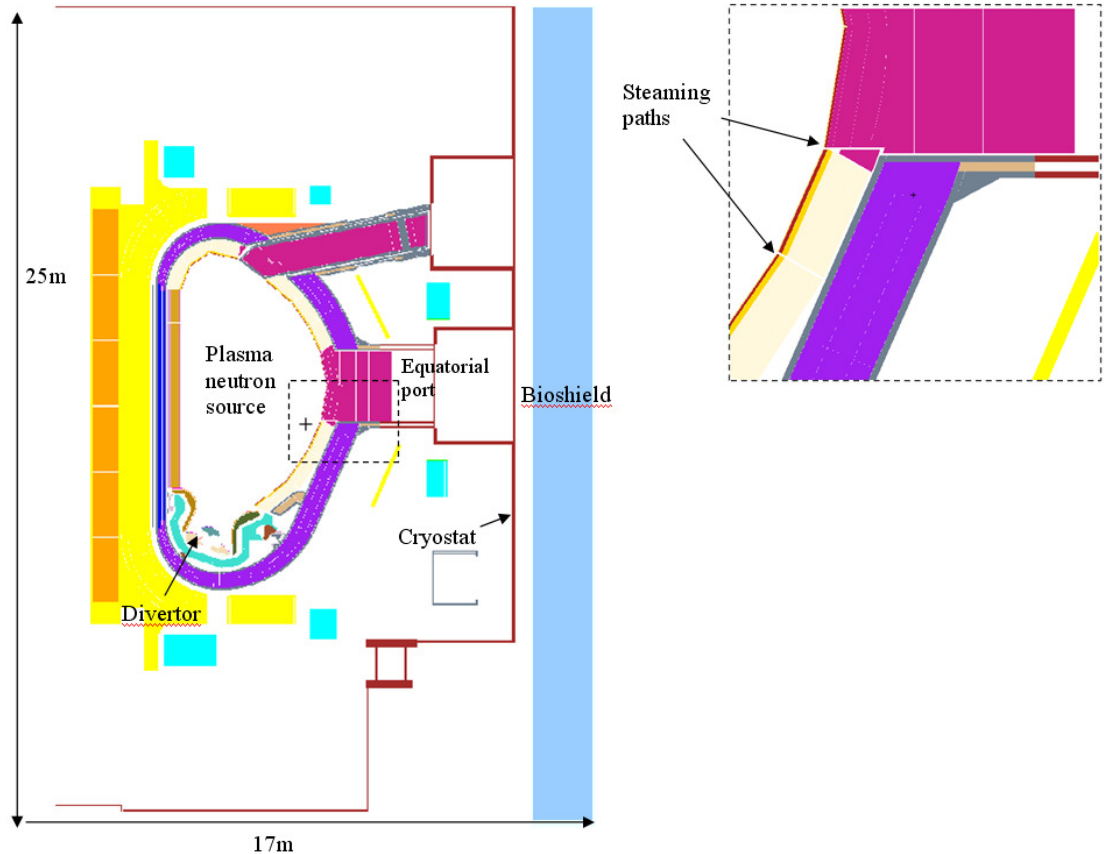

**Figure 1. Cross-section of ITER model 'A-lite'**

Where a streaming particle produces such a long history, the effect of streaming has clearly not been adequately captured in the WW generation process. An improved WW estimate with increased space and energy resolution, and the inclusion of particle direction (not a standard feature of the MCNP5), may alleviate this effect. However it is often impractical to obtain an 'ideal' WW map in a large and complex geometrical model, since memory limitations prevent the use of the high space and energy resolutions necessary to resolve streaming path effects. A practical solution to permit the use of such weight windows was therefore required.

### 3. LONG HISTORIES IN PARALLEL COMPUTATION

When an MCNP calculation running on a single machine encounters a LH, it is impossible to terminate the run until that history finishes, and the LH may run for hours or days (or longer). It is not possible to obtain tally results until the LH has been completed.

When running parallel calculations, the implications of LHs are more serious. Typically, MCNP tasks are distributed with equal batches of histories among available compute nodes. If one MCNP task encounters a LH which takes a disproportionate length of time to complete, the other tasks will perform no further work until the LH is completed, reducing both the available speedup from parallelisation, and the cluster utilisation. In addition, some computing clusters impose a maximum wall time, and clearly if the LH takes longer to complete than the allowed wall time, it will not be possible to obtain results before the job is terminated.

The impact of LH in parallel will also depend on the desired 'dumping cycle' of the calculation and the number of processors on which it runs. This effect was demonstrated using a simple stochastic analysis, for which it is assumed the long history problem can be represented as two discrete history types, 'normal' and 'long' (in reality, a continuous distribution of history lengths). For example, if a calculation runs 1000 normal histories per CPU-minute, and a LH takes 2000 CPU-minutes occurring randomly but on average every $10^7$ histories, then it can be shown that the maximum speed-up for such a case is approximately a factor of 10, and if run on 128 cores the cluster would be under 10% utilised (Figure 2). These parameters were chosen as an example but are typical of observations, as is the resulting parallel efficiency determined from this analysis. It is clear that long histories are highly detrimental to parallel computing.

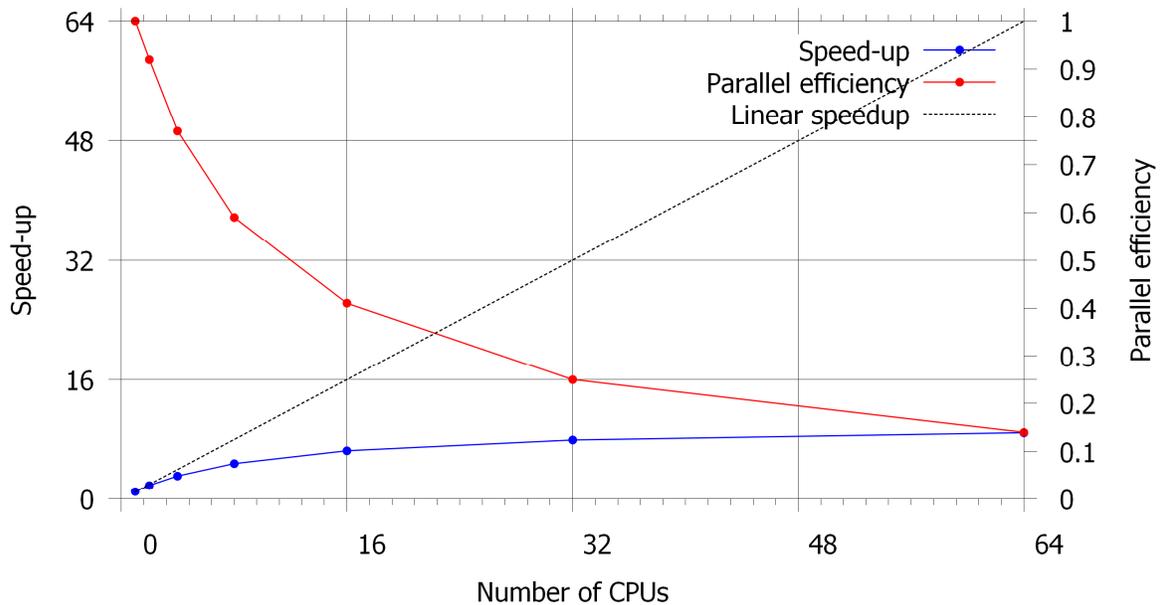

**Figure 2. Stochastic analysis results for example long history**

## 4. LONG HISTORY MITIGATION

It follows that for a Monte-Carlo calculation to run efficiently in parallel, either:
1. the LHs must be sampled sufficiently frequently, such that every CPU works on several LH during a dumping cycle; or
2. the length of the LH should be much less than the wall clock time of the dumping cycle of the calculation, so as to not significantly impede progress when one occurs.

The dumping cycle could be made longer than the typical long history duration, but this would not solve the issue of premature termination of a calculation due to a wall time limit. Source biasing could potentially be used to sample the streaming path more frequently; however this requires prior knowledge of the region that causes the LH. A practical solution would be to automatically reduce the length of the long histories when they do occur, and allow the history to complete in a reasonable time. CCFE has developed an adaptation to MCNP to perform this function.

The approach implemented was to renormalise the WW if, at any point in a history, the particle weight is more than a factor of M (a user-definable weight difference) above the weight of the region (cell or mesh voxel) defined by the WW. The entire WW is then renormalised such that the particle is only a factor of M higher than the WW, as shown in Figure 3.

Since the target weight range of the particle is now much closer to that of the particle weight, less splitting is performed. The revised WW map is used for the remainder of that particle history, after which it is reset for the next history. If the weight of a particle does not exceed the maximum weight difference threshold of 'M', then the WW is not adjusted. By varying the parameter 'M', the user can control the maximum allowed weight difference in a calculation.

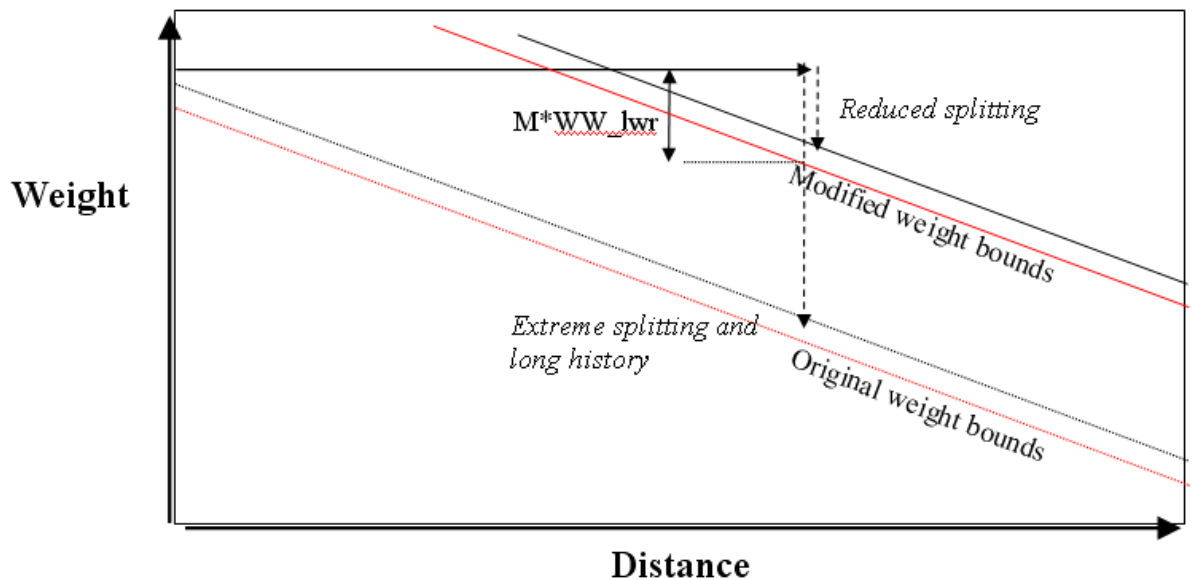

**Figure 3. Original and modified WW operation in the event of a LH**

## 5. TESTING AND RESULTS

### 2.1. Spherical Geometry

Testing was initially undertaken using a model which was intentionally designed to produce a long history. 14 MeV neutrons were born at the centre of a sphere of radius 200 cm, with material set to a steel and water mix. The sphere was divided into 20 spherical shells for variance reduction purposes, and a cell-based WW was produced to achieve accurate flux estimates throughout the sphere.

After constructing the WW, a conical penetration was added to provide a streaming path of known solid angle. The streaming path terminated in material so that the particle would be subsequently split by the WW. Since the streaming path was not accounted for in the production of the WW, high weight source particles will arrive with a weight in excess of the WW range, and lead to a long history. This artificial LH scenario is equivalent to a streaming path which was not adequately sampled during the production of the WW.

When particles were directed down this streaming path, they did indeed produce LHs, with the length of the LHs related to the depth of the penetration (and hence degree of weight difference). When the LH treatment was turned on, those histories completed in a shorter timescale, with smaller values of M producing shorter histories.

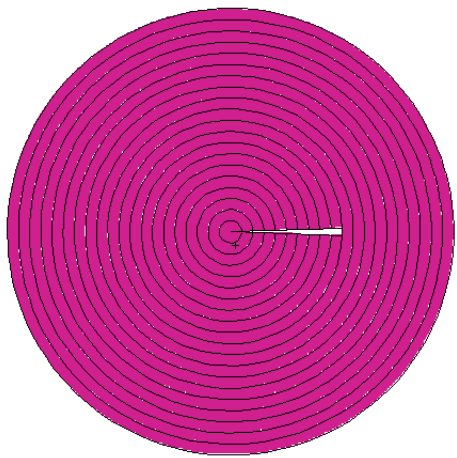

**Figure 4. Spherical test geometry with conical penetration**

Initial investigations, which will not be reported in detail here, focussed on the length of the streaming path and the resulting long histories with and without the WW treatment modification, with the aim of characterising an appropriate choice of the factor 'M' and relating it to the maximum long history length and parallel rendezvous cycle. Whilst a larger value of M permitted such histories to continue for longer timescales, the maximum permitted history length was found to also be dependent on the geometry, material, and WW resolution, and so no clear 'rule of thumb' could be established.

Further tests were undertaken to determine what effect the modified WW treatment had on the parallel efficiency of the calculation, and confirm the results were unchanged. The parallel efficiency is defined as the MCNP reported transport CPU time in minutes ('ctm' in the MCNP output file) divided by the product of the wall clock time and the number of CPUs requested ('computer time' in the output file). The angle of the streaming path was chosen such that approximately 1 in 4000 source particles passed down the penetration. The depth of the penetration corresponded to a weight difference of $\sim 5 \times 10^4$ between the WW and the source particle, with an average completion time of about 4 minutes. Clearly a 4 minute long history would be acceptable for most problems, but this scenario has been devised purely to demonstrate the effect, and also to be tractable when the long history mode is turned off.

The calculation was run for a range of values of 'M'; it was also run without the LH mode for comparison, and in analogue. It was found that the history rate varied as a function of 'M', and so the number of histories run was adjusted to keep the MCNP time CTM reasonably consistent. The results are shown in Table I.

**Table I: Results of LH mode on spherical test case**

| | | Analogue | $M=10^1$ | $M=10^2$ | $M=10^3$ | $M=10^4$ | $M=10^5$ | LH off |
|---|---|---|---|---|---|---|---|---|
| **Histories (NPS)** | | $10^8$ | $5 \times 10^6$ | $10^5$ | $10^5$ | $10^5$ | $10^5$ | $10^5$ |
| **MCNP time (CTM, min)** | | 423 | 322 | 394 | 396 | 407 | 461 | 450 |
| **Comp. time (CTME, min)** | | 446 | 341 | 489 | 492 | 639 | 1505 | 1465 |
| **Parallel efficiency** | | 95% | 94% | 80% | 81% | 64% | 31% | 31% |
| **Outer cell tally** | Mean | 4.08E-12 | 6.91E-12 | 4.53E-12 | 4.69E-12 | 4.68E-12 | 4.57E-12 | 4.57E-12 |
| | σ | 11% | 27% | 5.7% | 5.5% | 5.7% | 5.3% | 5.3% |

It is evident from Table I that the parallel efficiency is poor without the LH mode, and improves with the long history mode activated. As expected, for large values of M, the code reverts to the same behaviour as without the LH mode. The result is unchanged between the case without the LH mode, and with the LH mode activated (for $M \geq 100$). It is also observed that for very small values of M (e.g. 10), the statistical uncertainty on the tally increased, suggesting the weight shift was being invoked on most histories and is detrimental to the VR performance. The result for M=10 is higher than without the LH mode, however this is still consistent with the associated statistical uncertainty. For this particular scenario, values of 'M' of $10^2$ and $10^3$ produced acceptable behaviour in both VR performance and parallel acceleration.

Extremely low values of M are clearly not desirable, due to the increase in statistical uncertainty, however there was no apparent bias to the result. Analysis of the tally spectrum in 175 energy groups for M=1000 show results in excellent agreement with the standard WW treatment; flux bins over all energies lie well within an interval of 2σ around the mean (Figure 5) and show no biasing effects.

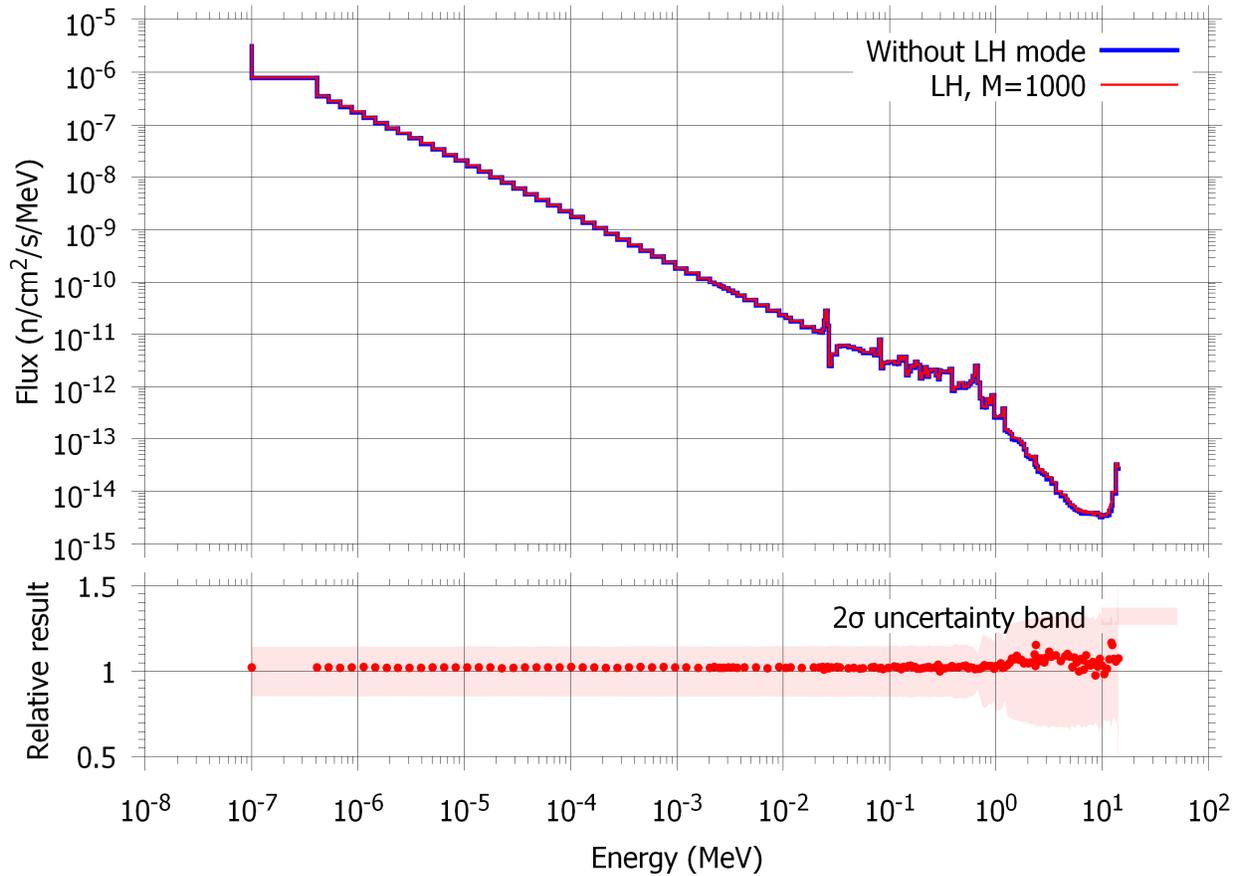

**Figure 5. Tally spectrum comparing LH/M=10³ to standard MCNP WW**

## 2.1. ITER A-lite Reference Model

The previous simple test proved that the proposed LH method works on a simple geometry and does not pollute tally results. Further testing was conducted on the ITER 'A-lite' model [7], which is the primary application of CCFE's GVR methods. The A-lite model contains many thousands of cells and surfaces, and as such requires parallel computation to make the problem tractable.

A previously generated GVR mesh weight window was utilised with the A-lite model. To visualise and quantify the effectiveness of the LH mode, a coarse 40 cm resolution mesh tally was used such that accurate statistics could be obtained in reasonable time. The model was run for various values of M, and the calculations were run to a similar MCNP reported run time 'ctm'. In order to remove any effect on tally standard deviation ($\sigma$) due to differences in MCNP run time, the Figure of Merit (FoM) was determined. The FoM is defined as:

$$FoM = \frac{1}{\sigma^2 \cdot time} \tag{1}$$

The FoM is therefore a number independent of computer time, which indicates the effectiveness of the GVR at reducing the statistical uncertainty (larger FoM is better). The tally standard deviation was taken to be the average over all mesh voxels (i.e. a 'global' FoM). Two values of FoM were reported, one based on the MCNP time ('ctm'), the traditional FoM reported by MCNP, and the other utilising the total computer time of the cluster. The latter therefore accounts for the parallel efficiency on the cluster, and is denoted pFOM ('parallel FoM').

It is expected that the LH method would decrease the FoM as the variance reduction is 'detuned' in order to make the calculation run more efficiently in parallel. The improvement in parallel efficiency must outweigh the reduction in efficiency for the method to be worthwhile - i.e. the method should give improvements in pFoM.

Results are given in Table II and Figure 6.

**Table II: Results of LH mode on ITER A-lite model**

|  | $M=10^3$ | $M=10^4$ | $M=10^5$ | $M=10^6$ | $M=10^7$ | $M=10^8$ | $M=10^9$ |
|---|---|---|---|---|---|---|---|
| **Histories run** | $4.5 \times 10^7$ | $1.5 \times 10^7$ | $1.0 \times 10^7$ | $1.0 \times 10^7$ | $1.0 \times 10^7$ | $1.0 \times 10^7$ | $1.0 \times 10^7$ |
| **CTM (mins)** | 22325 | 23651 | 23662 | 28028 | 31989 | 33447 | 43067 |
| **Comp. time (mins)** | 25400 | 30625 | 44850 | 64759 | 97114 | 111070 | 380852 |
| **Par. efficiency** | 88% | 77% | 53% | 43% | 33% | 30% | 11% |
| **Average Std Dev.** | 15% | 9.5% | 6.6% | 5.9% | 5.9% | 5.3% | 5.7% |
| **FoM** | 0.00199 | 0.00468 | 0.00984 | 0.01030 | 0.00906 | 0.01063 | 0.00719 |
| **pFoM** | 0.00175 | 0.00361 | 0.00519 | 0.00446 | 0.00298 | 0.00320 | 0.00081 |

When running the A-lite ITER model, it was found that without the LH mode, the problem was practically intractable, with a particular history taking over 5 days to finish. With the LH mode activated, the problem could be completed in a reasonable timeframe. The performance of the WW was found to degrade with low values of M, and particularly for $M \leq 10^4$. The parallel efficiency decreased with increasing M as expected, and for this particular model an optimum value of M that maximizes pFOM was found to be $10^5$ - $10^6$. Statistical error maps for $M=10^3$ and $10^5$ are shown in Figure 7, showing the increase in σ due to over-use of the LH mode.

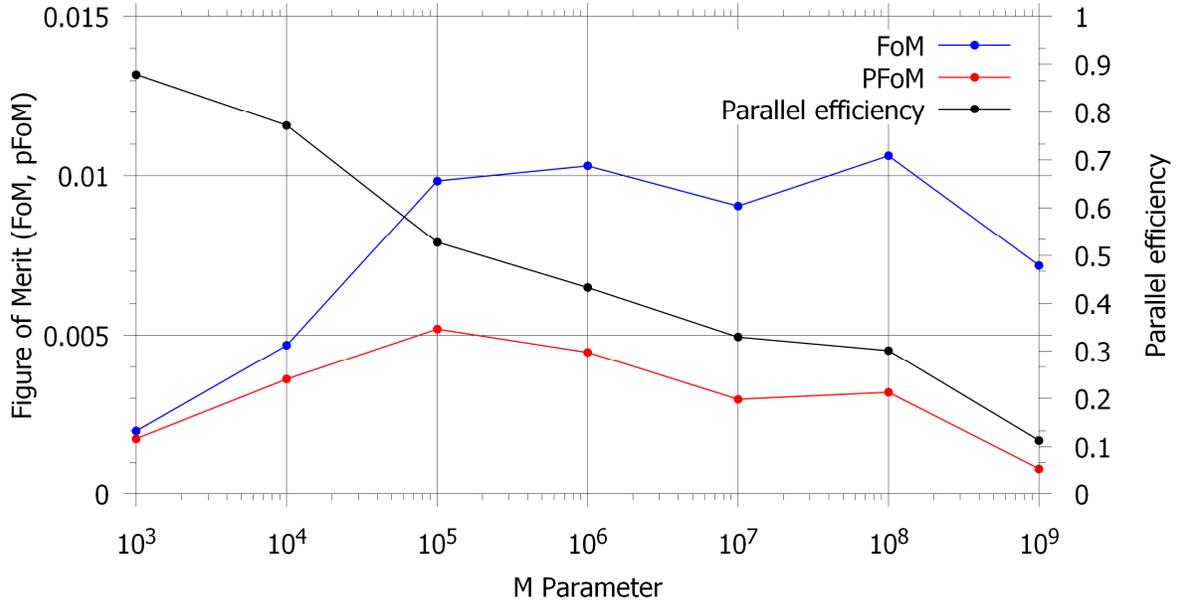

**Figure 6. FOM, pFOM and parallel efficiency v.s. M for the A-lite model**

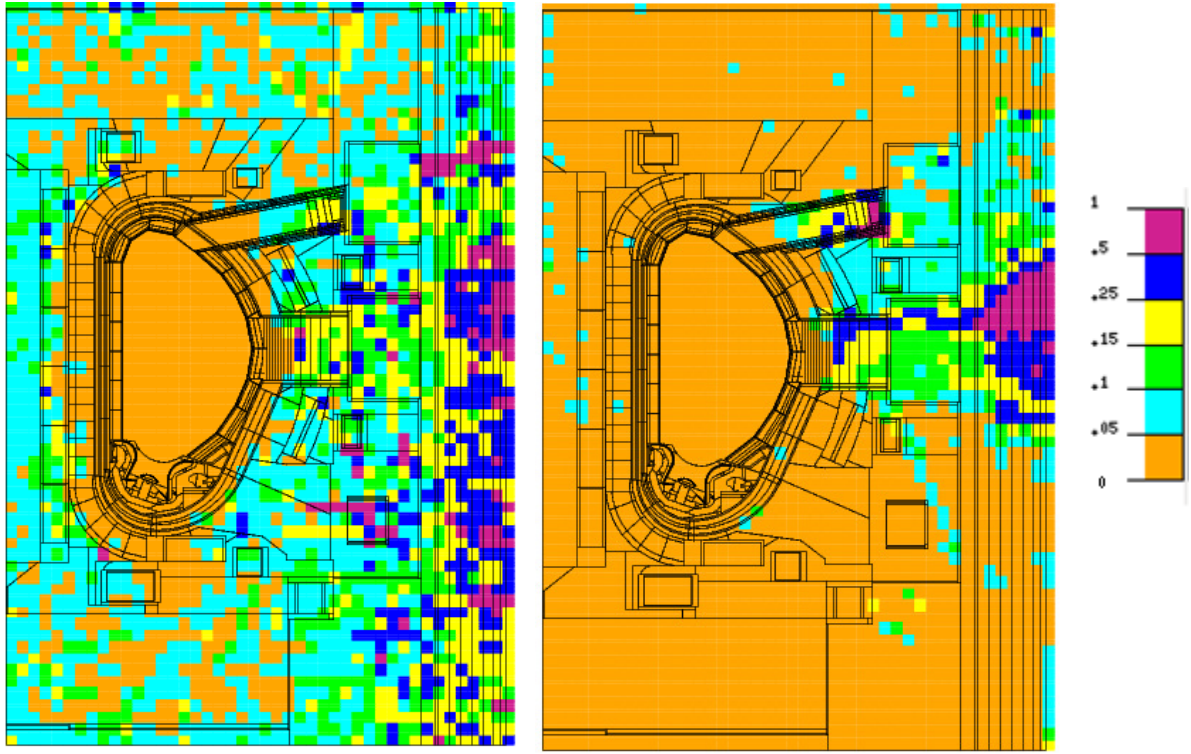

**Figure 7. Statistical error for M=$10^3$ (left) and M=$10^5$ (right)**

The use of the LH adaptation on the A-lite model demonstrates significantly improved effectiveness when running VR in parallel computations. By de-optimising the WW for troublesome histories, the reduction in VR effectiveness is more than offset by an improvement in parallel scale-up.

Since the calculations without the LH mode was not tractable, a direct comparison of the performance gain is not available, however the improvement in parallel FoM (error reducing power per unit cluster wall time) is at least a factor of 5 when comparing the $M=10^9$ to $M=10^5$ results, and would be greater than this when compared to the case without the LH modification.

### 3. CONCLUSIONS

Weight windows (WW) are used to perform variance reduction (VR) in MCNP. The WW is used to assign a target weight range for each discrete region in space and energy, in order to control the splitting and rouletting process and increase the sampling of important regions. The WW has a finite resolution in phase space (space and energy), and a target weight range is assigned is based on an average quantity (e.g. flux) in this phase space interval; however, infrequently sampled events can produce strong deviations from the average weight which in turn lead to excess splitting. This results in a history which takes a disproportionate amount of time to complete, a long history (LH), and the presence of this LH effect has been shown to reduce parallel cluster utilisation and in some cases make calculations intractable.

An example of a LH event would be a streaming path much smaller than the resolution of the WW mesh. A high weight source particle could enter such a streaming path, and be transported to an area of low expected weight, at which point it will undergo excessive splitting. In a case where a streaming particle is able to produce a long history, the effect of streaming has clearly not been adequately captured in the WW generation process, and a higher WW resolution may alleviate this effect. However it is this is often impractical to obtain an 'ideal' WW map in a large and complex geometrical model, since memory limitations prevent the use of the extremely high resolutions required. A more practical solution was therefore required.

CCFE has developed an adaptation of MCNP to counter the 'long history' problem. Where a particle is at a weight much greater than the expected range (by a user-adjustable parameter M), the weight window is modified for that history. The method allows calculations to proceed unhindered by long histories, improving the efficiency of parallel computing.

Testing has been conducted on a simple model, and the methodology has not been observed to bias the results, though the use of excessively small values for M has been shown to introduce higher statistical errors due to increased weight variation. Whilst it has been demonstrated that the method does not introduce bias to the mean, further work is required to examine the reliability of the tally statistical errors with regard to estimating convergence of the mean.

The use of the LH adaptation has been shown to reduce the effectiveness of VR (for small values of M), but also significantly improve the parallel speed-up of the MCNP calculation. By choosing appropriate value of M, the user can balance VR effectiveness against parallel efficiency, and obtain a net increase in performance. For the ITER A-lite model, an optimum of $M=10^5$-$10^6$ was found, which improved the parallel FOM by at least a factor of 5 compared to the standard WW treatment. In addition to improving computational efficiency, the method also prevents calculations from becoming intractable. The long history adaptation should prove beneficial to other geometrically complex models where streaming and deep penetration play an equally important role in the simulation.

## ACKNOWLEDGMENTS

This work was funded by the RCUK Energy Programme under grant EP/I501045 and the European Communities under the contract of Association between EURATOM and CCFE. The views and opinions expressed herein do not necessarily reflect those of the European Commission.